\documentstyle[aps,epsf,prl]{revtex}

\begin{document}
%%%%%%%%%enable the following command for two-column mode%%%%%%%%%%%%%%%%%%%%%
\twocolumn[\hsize\textwidth\columnwidth\hsize\csname@twocolumnfalse\endcsname

\title{Matrix Games, Mixed Strategies, and Statistical Mechanics}
\author{J.~Berg \thanks{email: johannes.berg@physik.uni-magdeburg.de} 
        and A.~Engel \thanks{email: andreas.engel@physik.uni-magdeburg.de}}
\address{Institute for Theoretical Physics\\
         Otto-von-Guericke University, Postfach 4120, 
         D-39016 Magdeburg, Germany
        }

\maketitle
 
\begin{abstract}
Matrix games constitute a fundamental problem of game 
theory and describe a situation of two players with completely conflicting 
interests. We show how methods from statistical mechanics can be used to 
investigate the statistical properties of optimal mixed 
strategies of large matrix games with random payoff matrices 
and derive analytical expressions for the value of the game 
and the distribution of strategy strengths. In particular the fraction of 
pure strategies not contributing to the optimal mixed strategy of a player 
is calculated. Both independently 
distributed as well as correlated elements of the payoff matrix are 
considered and the results compared with numerical simulations. 
\end{abstract}
\vskip.5pc]

\narrowtext

Game theory models in mathematical terms problems of strategic decision-making 
typically arising in economics, sociology, or international relations  
and owes much of its modern form to J. von Neumann \cite{NM}.
The generic situation in game theory consists of a set of players $\{X, Y,\ldots \}$ 
choosing between different {\it strategies} $\{X_i\},\{Y_i\},\ldots $, 
the combination of which determines the outcome of a game specified by the payoffs 
$P_X(X_i,Y_i,\ldots ), P_Y(X_i,Y_i,\ldots ), \ldots $ each player is going
to receive. 
The payoffs depend on the strategies of {\it all} players and
the problem for every individual player is to choose {\it his} strategy such
as to optimize his payoff without having control over the strategies of all
other players. Despite the extreme simplification of the real world
situation inherent in this framework, game theory has proven not
only to be a viable mathematical discipline but also to be able to 
characterize important features of economical systems. 
Many interesting results have been obtained since von Neumann's
pioneering work including the characterization of equilibria 
\cite{NM,Nash} and the emergence of cooperation \cite{Axelrod}. 
However, detailed investigations have been restricted 
either to general statements concerning e.g., the existence of equilibria, 
or to situations where every player has only a small number of strategies 
at his disposal and where the payoffs are simple functions of these strategies. 
As many situations of interest show a large number of possible 
strategies and rather complicated relationships between strategic 
choices and the resulting payoffs, 
it is tempting to model the payoffs by a random function and to apply the 
methods of statistical mechanics to describe the properties of the game. 
This will be a sensible approach if there are characteristic ``macroscopic'' 
quantities which do not depend on the particular realization of the 
random parameters, i.e. are {\it self-averaging} in the sense of 
the statistical mechanics of disordered systems \cite{MPV}, for 
related applications see \cite{greats}. 

In the present letter we show how methods from statistical mechanics 
can be applied to characterize the statistical properties of optimal 
strategies in matrix games with large randomly chosen payoff matrices. 
Explicitly we calculate the mean payoff and the fraction of pure strategies 
which occur in the optimal mixed strategy of a player.
For simplicity we restrict ourselves to 
matrix games, the type of zero-sum games between two players which also 
forms the basis of von Neumann's treatment \cite{NM,WJ}. 
Such games are defined by a 
(not necessarily square) payoff-matrix $c_{ij}$: 
Player $X$ may choose between $N$ strategies $X_i$ and player $Y$ 
between $M$ strategies $Y_j$ where $i=1,\ldots ,N$ and 
$j=1,\ldots,M$. At each step of this game they receive the 
payoffs $P_X(X_i,Y_j)=-P_Y(X_i,Y_j)=:c_{ij}$. 
As player $X$ wishes to gain as large a payoff $c_{ij}$ as possible, 
whereas player $Y$ must attempt to reach as small a value of $c_{ij}$ 
in order to maximize his payoff $P_Y(X_i,Y_j)=-c_{ij}$, the goals 
of the players are completely conflicting. 
Thus it is appropriate for the players to proceed as follows: 
Player $X$ knows that when playing strategy $X_i$ he will receive at 
least the payoff $\min_j c_{ij}$. He therefore chooses strategy $X_{i^*}$ 
satisfying $\min_j c_{i^*j}=\max_i\min_j c_{ij}$. Equivalently, 
player $Y$ plays strategy $Y_{j^*}$ determined by 
$\max_i c_{ij^*}=\min_j\max_i c_{ij}$ since it minimizes his losses 
for the optimal choices of $X$. It is easy to show that 
$\max_i\min_j c_{ij}\leq \min_j\max_i c_{ij}$ always. The situation is
simple if the matrix has a so-called {\it saddle-point}, i.e. 
if there is a pair $i^*,j^*$ satisfying
$\max_i\min_j c_{ij}=c_{i^*j^*}=\min_j\max_i c_{ij}$. In this case 
it is optimal for both players to stick to their {\it pure strategies} 
$X_{i^*}$ and $Y_{i^*}$ respectively, since deviations from an 
optimal strategy by one of the players will lead to a lower payoff 
for this player. For a large random matrix $c$ the probability
for the existence of a such a saddle point vanishes exponentially 
with the size of the matrix and the choice of an optimal strategy 
is less obvious. Since in this case 
$\max_i\min_j c_{ij}<\min_j\max_i c_{ij}$, player $X$ will attempt 
to achieve a greater gain than his guaranteed minimal gain 
$\max_i\min_j c_{ij}$ and likewise $Y$ will attempt to achieve a 
smaller loss than  $\min_j\max_i c_{ij}$. 
To this end they have to prevent their opponent from guessing 
which strategy they are going 
to play and choose each strategy with a certain probability 
$x_i$ and $y_j$ respectively\cite{NM}. A vector $x_i$ 
of probabilities is called a {\it mixed strategy} and 
by the normalization condition is constrained to lie on the 
$N$-dimensional simplex. The famous minimax theorem by von Neumann 
states that for any payoff matrix $c$ there exists a {\it saddle point 
of mixed strategies}, i.e. there are two vectors $x_i^*$ and 
$y_j^*$ such that 
\begin{displaymath}
  \max_{\{x_i\}}\min_{\{y_j\}} \sum_{ij} x_i\, c_{ij}\, y_j =
  \sum_{ij} x_i^*\, c_{ij}\, y_j^* = 
  \min_{\{y_j\}}\max_{\{x_i\}}\sum_{ij} x_i\, c_{ij}\, y_j \;.
\end{displaymath}
The expected payoff for the optimal mixed strategies
$\nu_c := \sum_{ij} x_i^*\, c_{ij}\, y_j^*$ is called the {\it value 
of the game} and $x_i^*$,$y_i^*$ denote optimal mixed strategies of 
player $X$ and $Y$ since again deviations from an optimal strategy 
by one of the players will lead to a lower payoff for this player. 

In the following we show how the statistical properties of such 
optimal mixed strategies for random payoff matrices may be 
characterized analytically in the limit $N \to \infty, M \to \infty$ 
with $M/N=\alpha=O(1)$.  As is generally 
the case in fully connected disordered systems, 
only the first two cumulants of the probability distribution 
$P(\{c_{ij}\})$ are relevant. Since an average value 
$\langle \langle c \rangle \rangle$ of the elements of the payoff 
matrix only results in a modified value of the game 
$\nu_c+\langle \langle c \rangle \rangle$ without changing the optimal 
mixed strategies, we may set $\langle \langle c \rangle \rangle=0$ 
without loss of generality and take the elements $c_{ij}$ to be 
independent Gaussian distributed variables with zero mean and 
variance $N^{-1}$. 

We then note \cite{WJ} that a necessary and 
sufficient condition for the mixed strategy $\{x_i\}$ of player 
$X$ to be optimal is 
\begin{equation}
  \label{opt}
  \sum_i x_i c_{ij} \geq \nu_c \  \forall j \ .
\end{equation}
The condition is necessary since if violated for some $j$ 
player $Y$ playing $Y_{j}$ will lead to a payoff lower than 
$\nu_c$. It is also sufficient since combining (\ref{opt}) with the 
minimax theorem gives $\sum_{ij} x_i\, c_{ij}\, y_j^*=\nu_c$. 
We may thus characterize mixed strategies of player 
$X$ by introducing the partition function 
\begin{equation}
  \label{part}
  Z(\nu)= \frac{
    \prod_{i=1}^N (\int_0^{\infty}dx_i) 
    \delta (\sum_{i=1}^N x_i-N)   \prod_{j=1}^{\alpha N} 
    \Theta (\sum_i x_i c_{ij}-\nu) }{
    \prod_{i=1}^N (\int_0^{\infty}dx_i) \delta (\sum_{i=1}^N x_i-N) }
  ,  
\end{equation}
where $\Theta (x)$ is the 
Heaviside step-function and the probabilities of playing a given 
strategy and the payoff have been rescaled so that 
$\sum_{i=1}^N x_i=N$ for convenience. 
Thus $Z(\nu)$ equals the fraction of the simplex obeying 
$\sum_i x_i c_{ij} \geq \nu \  \forall j$ and therefore lies on the 
interval $[0,1]$. 
Since $Z(\nu)$ scales exponentially with N, the quantity central to our 
calculation is the entropy $S(\nu):=1/N \ln Z(\nu)$, which
in general will be negative as usual for classical systems with 
continuous degrees of freedom. 

Assuming the entropy $S(\nu)$ to be self-averaging, 
we use the replica-trick $\ln Z = \lim_{n \to 0} \frac{d}{dn} Z^n$ 
and compute the average over the payoffs of the replicated partition 
function for integer $n$ ($a,b=1\ldots n$). The calculation 
proceeds by using the integral representation of the Heaviside 
step-function and by introducing the symmetric matrix of overlap 
order parameters $q_{ab}=1/N \sum_i x_i^a x_i^b$ via integrals over $q_{ab}$ 
and delta-functions represented by integrals over the conjugate order 
parameters $\hat{q}_{ab}$ \cite{Ga88}. The integrals over $E_a$ arise from
the integral representation of the constraint $\sum_i x^a_i=N$ giving 
\begin{eqnarray}
  \label{part2}
\lefteqn{\langle \langle Z^n(\nu) \rangle \rangle  = 
  \prod_{a \geq b} \int \frac{dq_{ab}d\hat{q}_{ab}}{2\pi/N} 
  \prod_{a} \int \frac{dE_{a}}{2\pi/N} } \hspace{0.5cm} \\ 
& & \exp(-iN\sum_{a \geq b}q_{ab}\hat{q}_{ab}-iN\sum_{a}E_{a}) \nonumber \\ 
& & \prod_{a,i} \int_0^{\infty} dx_i^a \exp (i \sum_{a \geq b,i}
   \hat{q}_{ab}x_i^ax_i^b
  +i\sum_{a,i}E_a x_i^a) \nonumber \\
& & \prod_{a,j} \int_{\nu}^{\infty} d\lambda_j^a \int 
  \frac{dy_j^a}{2\pi} 
  \exp(-\frac{1}{2}\sum_{a,b,j} q_{ab} y_j^a y_j^b +i\sum_{a,j}y_j^a 
  \lambda_j^a ) \nonumber 
\end{eqnarray}
In the limit of large payoff matrices $N \to \infty$ the integrals 
over order parameters are dominated by their saddle point. 
Throughout this paper we use the replica-symmetric ansatz \cite{rs_foot}

\begin{eqnarray}
  \label{rs}
  q_{aa}=q_1\qquad &    i\hat{q}_{aa}=-1/2 \, \hat{q}_1 \qquad &  iE_a=E \ \   
  \forall a \\
  q_{ab}=q_0 \qquad&   i\hat{q}_{ab}=\hat{q}_0 \qquad\qquad\; &   \forall a>b 
  \;.   \nonumber
\end{eqnarray}

The limit $n \to 0$ of (\ref{part2}) 
may now be taken by analytic continuation giving an entropy
\begin{eqnarray}
  \label{free}
\lefteqn{S(\nu)  =
  \text{extr}_{q_1,q_0,E,\hat{q}_1,\hat{q}_0}
   [\frac{1}{2}q_1 \hat{q}_1 + \frac{1}{2}q_0 \hat{q}_0 - E  
   +\frac{1}{2}\ln(2\pi)   } \hspace{.5cm} \nonumber\\ 
& &+\alpha \int Ds \ln H(\frac{\sqrt{q_0}s+\nu}{\sqrt{q_1-q_0}})
   -\frac{1}{2}\ln (\hat{q}_1+\hat{q}_0)-1  \qquad\quad \nonumber \\ 
& &+ \frac{\hat{q}_0+E^2}{2(\hat{q}_1+
  \hat{q}_0)} 
  +\int Dr \ln H(-\frac{\sqrt{\hat{q}_0}r+E}{\sqrt{\hat{q}_1+\hat{q}_0}})], 
\end{eqnarray}
where $Ds=\frac{ds}{\sqrt{2\pi}} \exp(-s^2/2)$, and 
$H(x)=\int_x^{\infty}Ds$. Numerical evaluation of (\ref{free}) shows that 
$S(\nu)$ is a continuously decreasing function of $\nu$. At $\nu_c$ it tends 
to $-\infty$, 
indicating that for larger values of $\nu$ there are no 
more solutions to (\ref{opt}). Furthermore as $\nu \to \nu_c$ one finds $q_0 \to q_1$ 
indicating that as the points contributing to (\ref{part}) crowd into 
an ever decreasing area of the simplex which shrinks to a point at $\nu_c$ 
their mutual overlap $q_0$ approaches the self-overlap $q_1$. 

In this regime the entropy may be 
conveniently written in terms of the order parameters 
$q_0,\hat{q}_0,E$, $\hat{w}=\hat{q}_1+\hat{q}_0$ and $v=q_1-q_0$. 
For $\nu<\nu_c$ $S(\nu)$ describes sub-optimal strategies. 
As $v \to 0$  we find 
$\hat{q}_0 \sim v^{-2}, \hat{w} \sim v^{-1}$.
Rescaling the conjugate order parameters accordingly and expanding 
the saddle-point equations to leading order in $v$ as $v \to 0$ we find 

\begin{eqnarray}
  \label{spe_a}
  \hat{w}-\alpha H(-\nu_c/\sqrt{q_0}) &=&0 \\
  \hat{w}-H(-E/\sqrt{\hat{q}_0}) &=&0 \nonumber \\
  \hat{q}_0-(\nu_c^2+q_0)\hat{w}-\alpha \sqrt{q_0} \nu_c G(-\nu_c/\sqrt{q_0}) 
  &=&0 \nonumber \\
  q_0-(E^2+\hat{q}_0)/\hat{w} - \sqrt{q_0} E/\hat{w}^2 G(-E/\sqrt{\hat{q}_0}) 
  &=&0 \nonumber
\end{eqnarray}   
with $E=q_0\hat{w}-\hat{q}_0$. 

The statistical properties of optimal strategies $\{x_i^{\text{opt}}\}$ 
may be deduced from the proportion of strategies $X_i$ with $x_i>a$
\begin{equation}
  \label{theta}
  \theta(a):=\langle\langle 1/N\sum_i\Theta(x_i^{\text{opt}}-a) \rangle\rangle 
  = H(\frac{\hat{w}a-E}{\sqrt{\hat{q}_0}}) \;. 
\end{equation}
Thus only a fraction $\theta(0)=\hat{w}$ of the pure strategies $X_i$ have 
$x_i>0$ and are played with non-zero probability. This striking effect may be explained by 
considering the behaviour of  player $Y$, whose optimal mixed strategy $y^*_j$ 
obeys $\lambda_i^*=\sum_j c_{ij} y_j^* \leq \nu_c \; \forall i$. 
Since $\nu_c= 1/N \sum_i x_i^* \lambda_i^*$, $x_i^*$ must be zero if 
$\lambda_i^*<\nu_c$. This mechanism thus ensures an expected payoff $\nu_c$ 
to $X$, even if $Y$ chooses an optimal strategy. However it is not to be 
confused with the concept of domination, widely discussed  in the game 
theory literature \cite{NM,WJ,Brandenburger}, where a strategy 
$X_i$ has $x_i=0$ because whatever the response of the opponent some 
other pure or mixed strategy will lead to a higher expected payoff. 
In fact in the thermodynamic limit domination of a pure strategy occurs with 
probability zero since for a pure strategy $X_k$ to be dominated by a mixed 
strategy $x^D_i$ requires $1/N \sum_i x^D_i c_{ij}\geq c_{kj} \forall j$ 
but the lhs is $O(N^{-1})$ whereas the rhs is $O(N^{-1/2})$. 

\begin{figure}[htb]
 \epsfysize=7.3cm
      \epsffile{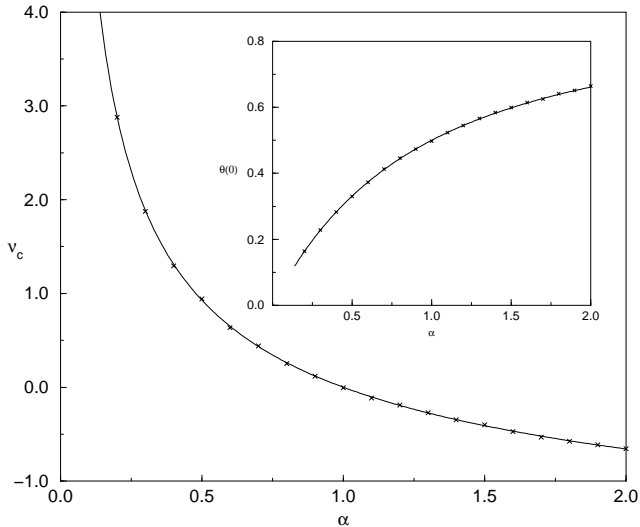}
\caption
{The value of the game $\nu_c$ and (inset) the fraction of 
strategies played with non-zero probability $\theta(0)$ as a function 
of $\alpha$. The analytical results 
(full line) are compared to numerical simulations with 
N=200 averaged over 200 samples. The symbol 
size corresponds to the statistical error.}   
\label{fig_a} 
\end{figure}  

Figure \ref{fig_a} shows the value of the game and (inset) 
the fraction of strategies played with non-zero probability 
as a function of the aspect ratio $\alpha$ of the payoff matrix. 
At $\alpha=1$ $\nu_c=0$ and $\theta(0)=1/2$.
The result $\nu_c=0$ at $\alpha=1$ is a consequence of the symmetry 
of the distribution of payoffs under $c_{ij} \rightarrow -c_{ji}$, 
i.e. under interchange of player $X$ and player $Y$\cite{nosym}.  
For $\alpha>1$ player $Y$ has a greater choice of strategies than 
player $X$ and vice versa. As expected, the payoff to player $X$ 
decreases as the range of strategy choices of player $Y$ increases. 
The fraction of strategies played with non-zero probability increases 
with $\alpha$, which reflects the decrease of $\nu_c$ with $\alpha$: 
At lower $\nu_c$ there are fewer $i$ with 
$\lambda_i^*=\sum_j c_{ij} y_j^* < \nu_c$, so as argued above the 
number of strategies $X_i$ played with non-zero probability 
increases as a result. 

\begin{figure}[htb]
 \epsfysize=7.3cm
      \epsffile{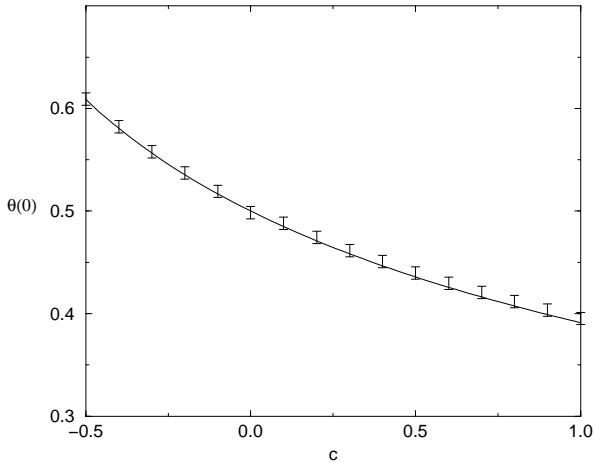}
\caption
{The fraction $\theta(0)$ strategies played with non-zero probability 
as a function of $c=c_r=c_c$.}
\label{fig_c}   
\end{figure}

We next abandon the initial assumption that the individual entries 
$c_{ij}$ in the payoff-matrix are independently distributed and 
consider the case where the outcomes 
of the game for different strategy choices of the players are 
correlated with each other. Such correlations may arise quite 
naturally in real applications since we expect some strategies 
to have broadly similar properties and hence yield similar results for
a given response of the respective opponent. For simplicity we restrict 
the discussion to the case $\alpha=1$. The most general tractable 
case appears to be
\begin{equation}
  \label{corr}
  \langle\langle c_{ij}c_{kl} \rangle \rangle / (
  \langle\langle c_{ij} \rangle \rangle 
  \langle\langle c_{kl} \rangle \rangle  )
  =:{\cal C}_{(ij)(kl)}
  ={\cal C}^c_{ik}{\cal C}^r_{jl}
\end{equation}
where ${\cal C}^c_{ik}$ and ${\cal C}^r_{jl}$ refer to column- and 
row-like correlations. Of course $P(\{c_{ij}\})$ is not uniquely 
determined by its second moments, but as argued above it suffices 
to consider Gaussian distributed payoff matrices so 
\begin{equation}
  \label{p_corr}
  P(\{c_{ij}\})=
  \frac{1}{\sqrt{(2\pi)^{N^2} \|{\cal C}\|}}
  \exp(-N/2 \sum_{ijkl}c_{ij} {\cal C}^{-1}\,_{(ij)(kl)}c_{kl})
\end{equation}
The specific form of ${\cal C}^c_{ik}$ and ${\cal C}^r_{jl}$
we will consider in the following is 
\begin{equation}
  \label{c_form}
  {\cal C}^{c,r}_{ik}=\left\{ \begin{array}{ll} 1 &  \;\;i=k    \\
                                        c_{c,r}/N   & \;\; i\neq k\\
                               \end{array} \right.     
\end{equation}
and the resulting replica symmetric entropy averaged over the 
distribution (\ref{p_corr}) may be calculated as outlined for the 
case of uncorrelated payoffs above. Again in the limit $v \to 0$ 
the corresponding 
saddle point equations describe optimal strategies. For 
$c_c=c_r=c$ the optimal payoff is zero as a result of the 
symmetry of (\ref{p_corr}) under the exchange of players. 
Figure \ref{fig_c}\ shows the fraction $\theta(0)$ of strategies 
played with non-zero probability in optimal strategies as a 
function of $c$. $\theta(0)$ decreases with 
increasing $c$: At positive $c$ there are strategies which tend to 
be beneficial for player $X$ whatever the response of the opponent. 
As a result $X$ concentrates on a smaller fraction of his strategies 
and vice versa for negative $c$. 

\begin{figure}[htb]
 \epsfysize=7.3cm
      \epsffile{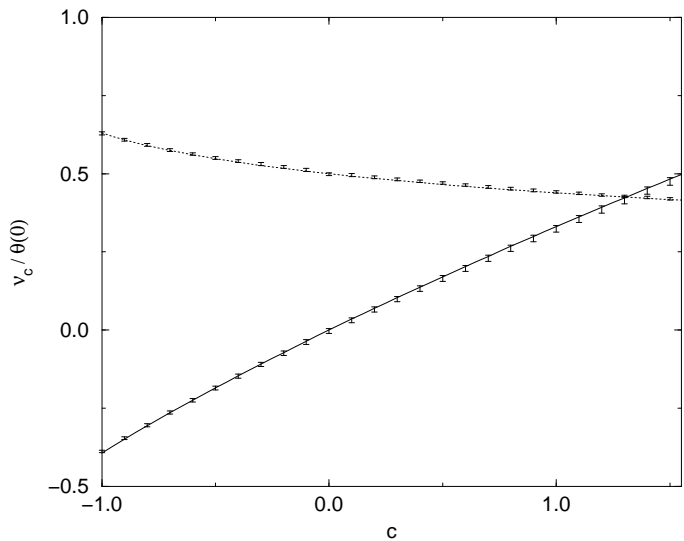}
\caption
{The optimal payoff $\nu_c$ (full) and the fraction $\theta(0)$ 
of strategies played with non-zero probability (dotted) against 
$c=c_r$ at $c_c=0$.}  
\label{fig_cr} 
\end{figure}

In the asymmetric case $c_r=c, c_c=0$ however a non-zero value of the game is 
possible. 
The resulting value for $\nu_c$ and the fraction of strategies played with 
non-zero probability are shown in figure \ref{fig_cr}. 
Again the correlation between payoffs in the same row of the payoff 
matrix lead to strategies which tend to be 
either beneficial or detrimental to player $X$. 

By admitting only the 
beneficial ones into his mixed strategies, $X$ may achieve a 
positive payoff. The fraction of strategies played with non-zero 
probability decreases accordingly. For negative $c$ the entries in 
the same rows of the payoff matrix are anticorrelated, so different 
responses of player $Y$ to the same strategy 
of player $X$ tend to lead to different payoffs. This 
situation leads to a negative value of the game.  
The simulation results shown in figures 1-3 were obtained using the 
simplex-algorithm to solve the linear programming 
problem \cite{WJ} defined by (\ref{opt}) for a system of size $N=200$ 
averaged over 200 payoff matrices with Gaussian distributed 
inputs\cite{ising}.The numerical results show very good agreement 
with the analytical expressions.    

In conclusion we have shown that techniques from the statistical 
mechanics of disordered systems may be used to analyze the 
statistical properties of optimal solutions of matrix games with 
random payoffs. Self-averaging ``macroscopic'' quantities 
such as the value of the game were identified and calculated for 
various probability distributions. These quantities include the 
fraction of strategies played with non-zero probability.     
Further problems in matrix games which may be treated using these methods 
include the effects of deviating from the optimal strategy and the influence of 
perturbations of the payoff matrix on the optimal strategy, 
which forms the basis of the justification for the full stability of 
mixed equilibria\cite{Harsanyi}. Furthermore work is in progress on 
a field of interest to current mathematical game theory, the 
statistical description of Nash-equilibria in bimatrix games.

{\bf Acknowledgments}: It is a pleasure to thank R. Monasson, L. Schneider, 
and M. Weigt for stimulating discussions. JB gratefully acknowledges financial 
support by the \it{Studien\-stiftung des Deutschen Volkes}. 

\bibliographystyle{unsrt}

\begin{thebibliography}{99}
\bibitem{rs_foot} Replica symmetry may be justified from the convexity 
of the solution space of equation (\ref{opt}), furthermore it has been 
found to be locally stable for $\nu < \nu_c$.
\bibitem{nosym} Under $\alpha \rightarrow 1/\alpha$ 
$\nu_c \rightarrow -\sqrt{\alpha} \nu_c$ since the variance 
of the payoffs is $N^{-1}$.
\bibitem{ising} Virtually identical results have been obtained for different 
probability ensembles such as $c_{ij}=\pm N^{-1/2}$.
\end{thebibliography}

\end{document}